# Optical Magnetic Mirrors without Metals

Sheng Liu,[1,2,*] Michael B. Sinclair,[1] Thomas S. Mahony,[1,2] Young Chul Jun,[1,2] Salvatore Campione,[3] James Ginn,[1] Daniel A. Bender,[1] Joel R. Wendt,[1] Jon F. Ihlefeld,[1] Paul G. Clem,[1] Jeremy B. Wright,[1] and Igal Brener[1,2,*]

[1]*Sandia National Laboratories, Albuquerque, NM 87185, USA*
[2]*Center for Integrated Nanotechnologies, Sandia National Laboratories, Albuquerque, NM 87185, USA*
[3]*Department of Electrical Engineering and Computer Science, University of California Irvine, Irvine, CA 92697, USA*
[*]*e-mail: ibrener@sandia.gov, snliu@sandia.gov*

**The reflection of an optical wave from a metal, arising from strong interactions between the optical electric field and the free carriers of the metal, is accompanied by a phase reversal of the reflected electric field. A far less common route to achieve high reflectivity exploits strong interactions between the material and the optical magnetic field to produce a "*magnetic mirror*" which does not reverse the phase of the reflected electric field. At optical frequencies, the magnetic properties required for strong interaction can only be achieved through the use of artificially tailored materials. Here we experimentally demonstrate, for the first time, the magnetic mirror behavior of a low-loss, all-dielectric metasurface at infrared optical frequencies through direct measurements of the phase and amplitude of the reflected optical wave. The enhanced absorption and emission of transverse electric dipoles placed very close to magnetic mirrors can lead to exciting new advances in sensors, photodetectors, and light sources.**

Magnetic mirrors or high impedance surfaces were first proposed at microwave frequencies [1]. An important advantage of these mirrors is that a transverse electric dipole placed close to the mirror surface is located at an antinode of the total (incident plus reflected) electric field and, hence, can absorb and emit efficiently[2]. In contrast, a dipole placed close to a metal surface experiences a node of the total electric field and can neither absorb nor emit efficiently. At microwave frequencies these exceptional properties of magnetic mirrors have been utilized for smaller, more efficient antennas and circuits[3-6]. Magnetic mirrors can also exhibit unusual behavior in the far-field, through the appearance of a "magnetic Brewster's angle" at which the reflection of an s-polarized wave vanishes[7,8].

At optical frequencies, magnetic behavior can only be achieved through the use of artificially tailored materials and, as a result, relatively little work on optical frequency magnetic mirrors has been reported thus far. Recent investigations of magnetic mirror behavior at optical frequencies have utilized metallic structures such as fish-scale structures[9] and gold-capped carbon nanotubes[10]. However, the metals utilized in these approaches suffer from high intrinsic Ohmic losses at optical frequencies. All-dielectric metamaterials, based upon subwavelength resonators, with much lower optical losses and isotropic optical response have been used to demonstrate fascinating properties in a number of recent investigations[11-25]. In another recent work, magnetic mirror behavior was theoretically predicted for silicon dielectric resonators in the near infrared[26]. Although the reflection amplitude spectrum was measured in this work, no experimental phase measurements were achieved. In principle, this work is the same as our previous work[12] that only showed high reflectivity at the magnetic dipole resonance, which is a necessary but not sufficient condition for magnetic mirror behavior. Moreover, the use of silicon as a resonator material does not result in a sufficiently small array spacing for effective medium behavior, particularly at oblique angles. The benefit of using higher refractive index materials (Tellurium (Te) in our work) is clear by comparing the reflectivity amplitude spectra of the silicon structure of ref. [26] and the Te structure in this work: the Te resonators exhibit a much better spectral separation of the electric and magnetic dipole resonances. Recently theoretical work[27,28] has also predicted nearly total omnidirectional reflection based upon the interplay between different resonant modes of high refractive index resonators.

This is in strong contrast to the magnetic mirror behavior obtained in the effective medium limit which does not depend on the coupling between different resonant modes. Therefore, we present here the first experimental demonstration of optical magnetic mirror using an all-dielectric metamaterial. Furthermore, none of these previous optical magnetic mirror (OMM) demonstrations, including works that utilized metallic structures, were able to provide detailed temporal information about the optical fields such as the resonant build-up of the response after transient excitation. In the present work, we overcome these limitations by using an OMM based on a sub-wavelength two-dimensional array of dielectric resonators fabricated from a low-loss, high permittivity dielectric material: Te. Furthermore we utilize a phase-sensitive, time-resolved optical technique to provide direct experimental proof of the magnetic mirror behavior. We also show that the electric field standing wave pattern for plane wave illumination exhibits an antinode at the surface of the OMM, indicating that efficient coupling to transverse electric dipoles placed close to the mirror surface should be possible.

Fig. 1a shows a schematic of our magnetic mirror which comprises a two-dimensional array of sub-wavelength, Te cube resonators. The lowest frequency resonance of a high permittivity cubic resonator exhibits a circular displacement current pattern which yields magnetic dipole behavior[12,29], while the next higher resonance leads to a linear displacement current and, hence, electric dipole behavior. The large permittivity of Te[12,29,30] ensures that the dimensions of the resonator and array spacing are sufficiently sub-wavelength. Two-dimensional arrays of Te resonators were fabricated by depositing Te on a $BaF_2$ substrate, followed by e-beam lithography patterning and reactive ion etching. The resulting cube-like resonators exhibited a height of 1.7 μm and a base of ~1.5 × 1.5 μm (a slight over-etching led to the deviation from perfect cube geometry). $BaF_2$ was selected as a substrate material due to its low reflective index and low loss throughout the IR spectral region. The large refractive index contrast between Te and $BaF_2$ allows for a high degree of confinement within the Te resonator and small leakage into the substrate[17]. The unit-cell spacing of the array was 3.4 μm for a ~45% duty cycle[12]. Fig. 1b shows a scanning electron microscope (SEM) image of the fabricated OMM sample. Fig. 1c shows the measured reflection spectrum of the sample which exhibits two reflection maxima that are close to the magnetic and electric dipole resonances (the electric field patterns are shown in Fig. 1c). The wavelengths of the magnetic (8.95 µm) and electric (7.08 µm) dipole resonances do not precisely correspond to the transmission minima[31], but rather are determined by the loss maxima (i.e. absorption maxima) (see Supplementary Information S1). This selection is also supported by numerical simulations, which show that the strongest dipole field intensities occur close to the wavelengths of maximum absorption (see Supplementary Information S1).

To directly measure both the amplitude and phase of the electric field of an optical wave reflected from the OMM, we utilize phase-locked time-domain spectroscopy (TDS)[32]. TDS has proven to be a powerful technique at terahertz frequencies[33-36] and has been recently extended to higher mid-infrared (mid-IR) frequencies[37,38] — a highly interesting frequency region covering vibrational and electronic resonances of molecular system and solids[39,40]. Fig. 2 shows a simplified schematic of our stable-carrier-envelope phase (CEP) locked mid-IR TDS system. Briefly, an ultrafast fiber laser system was used to generate ~250 femtosecond (*fs*) mid-IR pulses, tunable between 8.1 µm and 11 µm. The p-polarized pulses were focused onto either the OMM sample or a gold reference surface (which was deposited on top of the Te in an unpatterned region of the sample) at an incidence angle of 30 degree with ~10 degrees of angular divergence in the focused beam. Thus, the incident radiation covered a range of angles from 25-35 degrees. Care was taken to ensure that switching between the gold and OMM surfaces did not cause any spurious delay change (see Supplementary Information S2). A synchronized 15 fs gate pulse output from the same fiber laser was used to measure the reflected infrared electric-field transients through phase-matched electro-optic sampling in a GaSe crystal[32,41]. More details of our TDS system can be found in Supplementary Information S3, ref. 19 and 42.

Fig. 3a shows the measured electric field of the reflected pulses from the OMM (red line) and gold surface (blue line) when the central frequency of the incident pulse coincides with the magnetic dipole resonance of the Te resonators at 8.95 µm. For comparison, Fig. 3b is the simulated reflected electric field obtained using a commercial finite-difference time-domain (FDTD) simulator, showing remarkable agreement between the experimental and simulation data. For both experiment and simulation, the phase of the reflection from the gold surface was referenced to a plane very close to the center of the cubic resonators, which is the plane that contains the radiating dipoles (see Supplementary Information S4). Several features are noticeable in both the measured and simulated waveforms in Fig. 3a-b. First, the amplitude of the red curve is slightly smaller than that of the blue curve, indicating that the reflectivity of the OMM is slightly smaller than that of the gold surface (which is close to a perfect mirror in the mid-IR). Second, the reflected field envelope from the OMM has a ~50 *f*s delay (approximately two optical periods) compared to that from the gold surface due to the resonant interaction of the optical pulses with the metasurface resonators. Third, and most importantly, the electric field fringes from the OMM are nearly phase-reversed with respect to the fringes from the gold reference (which are phase-reversed with respect to the incident field). Thus, the electric field waveforms of Fig. 3a demonstrate that the field reflected from the OMM is in-phase with the incident field, which, combined with the high reflection amplitude, directly demonstrates magnetic mirror behavior at wavelengths around 9 µm. Due to the limited wavelength tuning range of the IR pulses, we further studied the phase shift at the electric dipole resonance using FDTD simulations only (see Supplementary Information S5). These simulations show that the electric field fringes from the OMM are in-phase with those from the gold surface at the electric dipole resonance, because the OMM acts as a normal mirror at this electric resonance frequency.

For a quantitative analysis of the phase difference between the electric fields reflected from gold and the OMM, we performed a Fourier transform of the measured field transients (Fig. 3c). The time harmonic convention $\exp(-i\omega t)$ is implicitly assumed throughout the manuscript. To cover a broader spectral range, a second data set was obtained with the center frequency of the incident mid-IR pulses tuned to ~8.1 µm. The electric field phase obtained from the FDTD simulation results is also plotted in Fig. 3c (blue curve) for comparison. Although the experimental data (red and blue circles) does not reach the electric dipole resonance frequency, it agrees very well with simulations over a wide wavelength region between 7.6 µm and 9.3 µm. We observe a ~180 degree phase difference between the magnetic and electric dipole resonances, further confirming the magnetic mirror behavior near 9 µm. In addition, we measured another OMM sample made with a different fabrication technique resulting in a smaller spectral separation between the electric and magnetic dipole resonances. For this sample, our mid-IR pulses are spectrally broad enough to cover both resonances at once (see Supplementary Information S6). Once again, we observe an approximately ~180 degree phase difference between the two dipole resonances which further demonstrates the OMM behavior at the magnetic dipole resonance. The fact that the behavior does not sensitively depend upon the separation of the electric dipole and magnetic dipole demonstrates that the magnetic mirror behavior does not rely on the coupling between these resonances.

To analyze the magnetic mirror behavior, we model each cubic resonator as an electric dipole and a magnetic dipole (whose polarizabilities are calculated via full-wave simulations) located at the center of the cube. We compute the amplitude and phase of the reflection coefficient of the 2D array including cross coupling between the electric and magnetic dipoles (see Supplementary Information S4) using two-dimensional periodic dyadic Green's functions[43,44,45] and obtain very good agreement with full-wave simulations. Near the magnetic resonance the magnetic dipole behavior dominates the optical response and magnetic mirror behavior is observed. In contrast, near the electric resonance the electric dipole dominates and the array behaves like an electric mirror with an electric field phase shift of 180 degrees upon reflection.

When a transverse electric dipole is placed in the near-field of a conventional electric mirror, its emission is largely canceled by that of its image dipole[2,46-49]. In contrast, the image dipole produced by a magnetic mirror is in-phase with the original dipole and emission is allowed. Similarly, a transverse electric dipole in very close proximity to an electric mirror finds itself at the node of the total electric field under plane wave illumination and cannot efficiently absorb incoming radiation, whereas an electric dipole placed near a magnetic mirror is located at the antinode of the total electric field and can absorb efficiently. As a result, optical magnetic mirror behavior has been studied extensively at microwave frequencies for efficient, compact microwave circuits and antennas[3-6]. To demonstrate that these advantages can also be obtained at optical frequencies, we utilized FDTD simulations to generate maps of the total electric field at both the electric and magnetic resonance wavelengths of the OMM and compared them to the total field maps obtained for a conventional gold mirror. Fig. 4a shows the standing wave patterns obtained at the electric resonance wavelength of the OMM for both the OMM and a gold surface. As discussed previously, the gold surface is located at a height equivalent to the center of the dielectric resonators. At electric resonance, the two standing wave patterns show similar behavior, with an electric field node located at the surface of both mirrors. The behavior is distinctly different at the OMM magnetic resonance wavelength (Fig. 4b), where an electric field antinode is observed at the surface of the OMM. As expected, the gold mirror still exhibits a surface node at this wavelength. (Supplementary Movie shows a movie of a plane wave reflected from a gold surface and an OMM are in-phase and our-of-phase at the magnetic and electric resonances, respectively.) Closer inspection of Fig. 4b shows that the total electric field exhibits some enhancement at the mirror surface, which is presumably due to the resonant nature of the OMM. Hence, at this wavelength, dipole absorbers/emitters placed in the immediate vicinity of the OMM surface are expected to interact strongly with the total field and efficiently absorb/emit electromagnetic energy. Indeed, we observe a large enhancement of dipole radiative emission when a transverse electric dipole at the magnetic resonance wavelength is placed close to the OMM as shown in Fig. 5. Using full wave simulation tools, the radiative decay rate is calculated by measuring the outgoing flux from a dipole source and normalizing it by the flux in free-space. We calculate the normalized radiative decay rate of an electric dipole oscillating at the magnetic resonance frequency as a function of the dipole-surface separation for both a 5 x 5 resonator array approximating our OMM and a gold surface. The behavior for these two cases is strikingly different: i) the oscillatory dependence on distance is shifted by about half a period, which is a further confirmation of the magnetic mirror behavior of the OMM; ii) while the emission from the dipole is quenched very close to the gold surface, the dipole emission rate near the magnetic mirror is enhanced even for very small distances. However, the emission from the transverse dipole in the near-field of the OMM is more complicated than what is expected from simple image dipole arguments, possibly due to coupling to higher order modes of the 5 x 5 array.

Thus, we conclude that all-dielectric magnetic mirrors are good candidates for new types of infrared sensors such as remotely interrogated chemical sensors in which molecular species adsorbed onto chemically selective layers on the mirror surface will interact strongly with the interrogating radiation and impart spectral signatures on the reflected beam. In addition, these mirrors will be useful for compact and efficient thermal radiation sources in which thermal emitters are deposited directly on the mirror surface and the emitted thermal energy is efficiently coupled to the far-field. Moreover, due to the zero phase shift of the electric field upon reflection, it should be possible to construct a $\lambda/4$ optical cavity bounded by a magnetic mirror on one side and an electric mirror on the other. Although much of the interest in magnetic mirrors stems from their near-field behavior, magnetic mirrors can also exhibit unusual behavior in the far-field. In particular, such mirrors can exhibit a "magnetic Brewster's angle" at which the reflection of an s-polarized wave vanishes[7,8], leading to new types of polarization control devices. Further, we envision that through appropriate metasurface design it should be possible to produce an admixture of magnetic mirror and electric mirror behaviors to achieve a pre-desired angular response for the amplitude and phase of reflected waves.

.

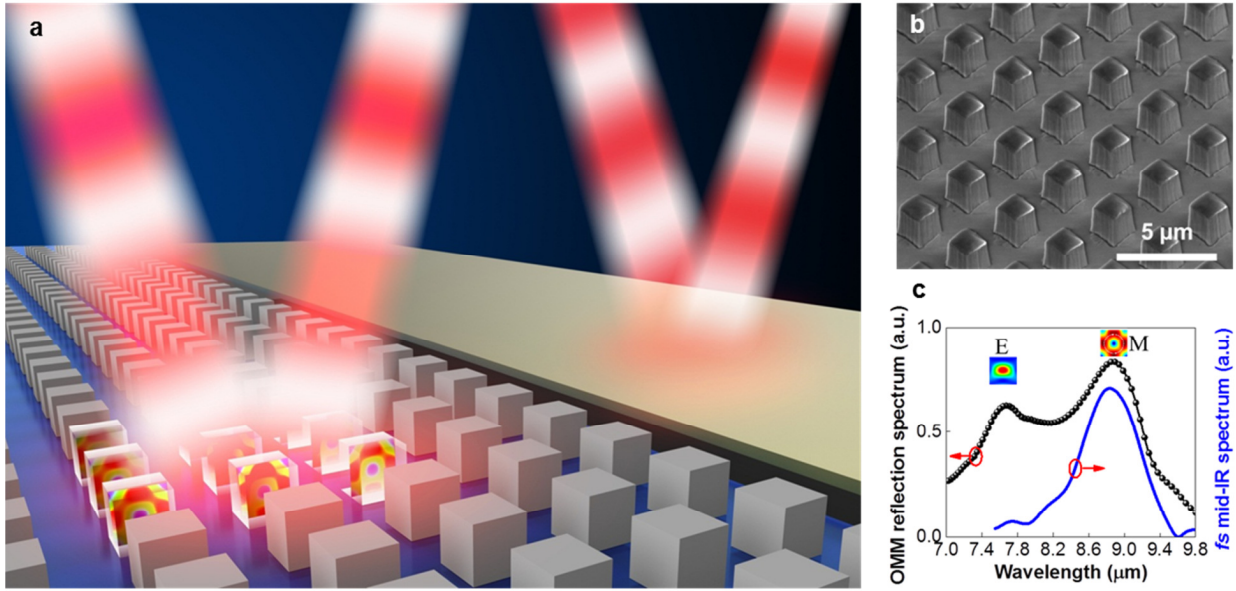

**Figure 1 | Principle of all-dielectric optical magnetic mirrors. a**, The cubic dielectric resonators on the left side do not induce a phase shift of the reflected electric field at the magnetic resonance, but rather act as a dielectric magnetic mirror in the optical frequency range. On the contrary, the gold surface on the right side (which serves as a reference surface) exhibits a 180 degree phase shift of the electric field upon reflection. **b**, SEM image of the Te cube dielectric metasurface of our OMM. **c**, The reflection spectrum of the metamaterial sample in **b** shows two reflection maxima corresponding to the lowest (magnetic dipole) and the second lowest (electric dipole) resonances.

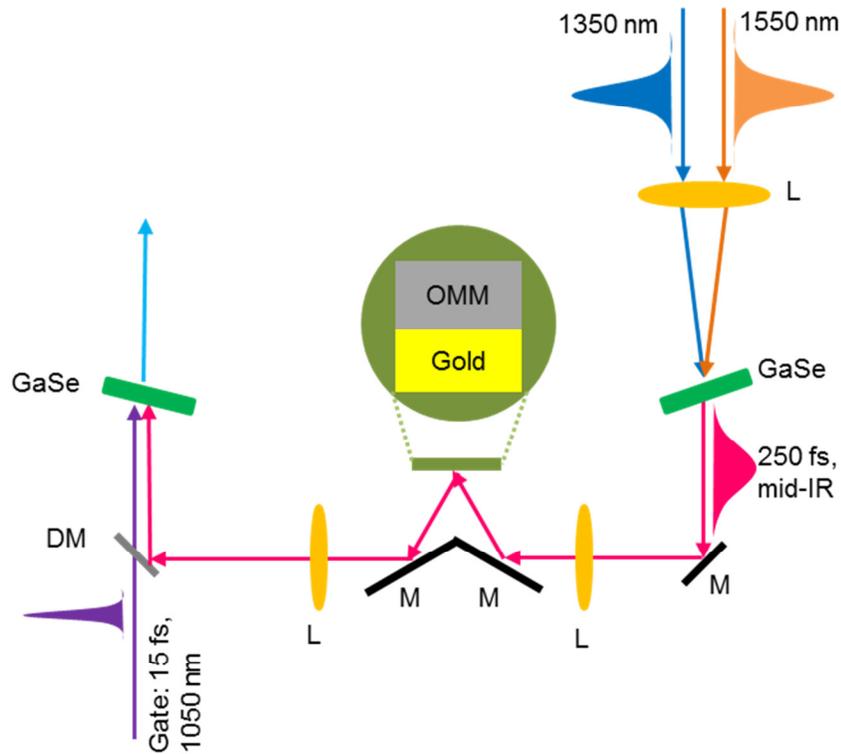

**Figure 2 | Schematic of the time-domain spectroscopy (TDS) set-up.** We used a phase-locked TDS in the mid-IR to directly measure the phase shift of the reflected electric field from the OMM. Mid-IR pulses of 250 fs duration, produced by difference frequency mixing between ~1.35 µm and ~1.55 µm pulses, were focused by a ZnSe lens onto the OMM or gold surface. Another ZnSe lens was used to collect the reflected mid-IR beam. Gate pulses with a wavelength of 1.05 µm and duration of 15 fs were combined with the mid-IR pulses using a dichroic mirror and then focused into another GaSe crystal for phase matched electro-optic sampling. DM, dichroic mirror; L, lens; M, mirror.

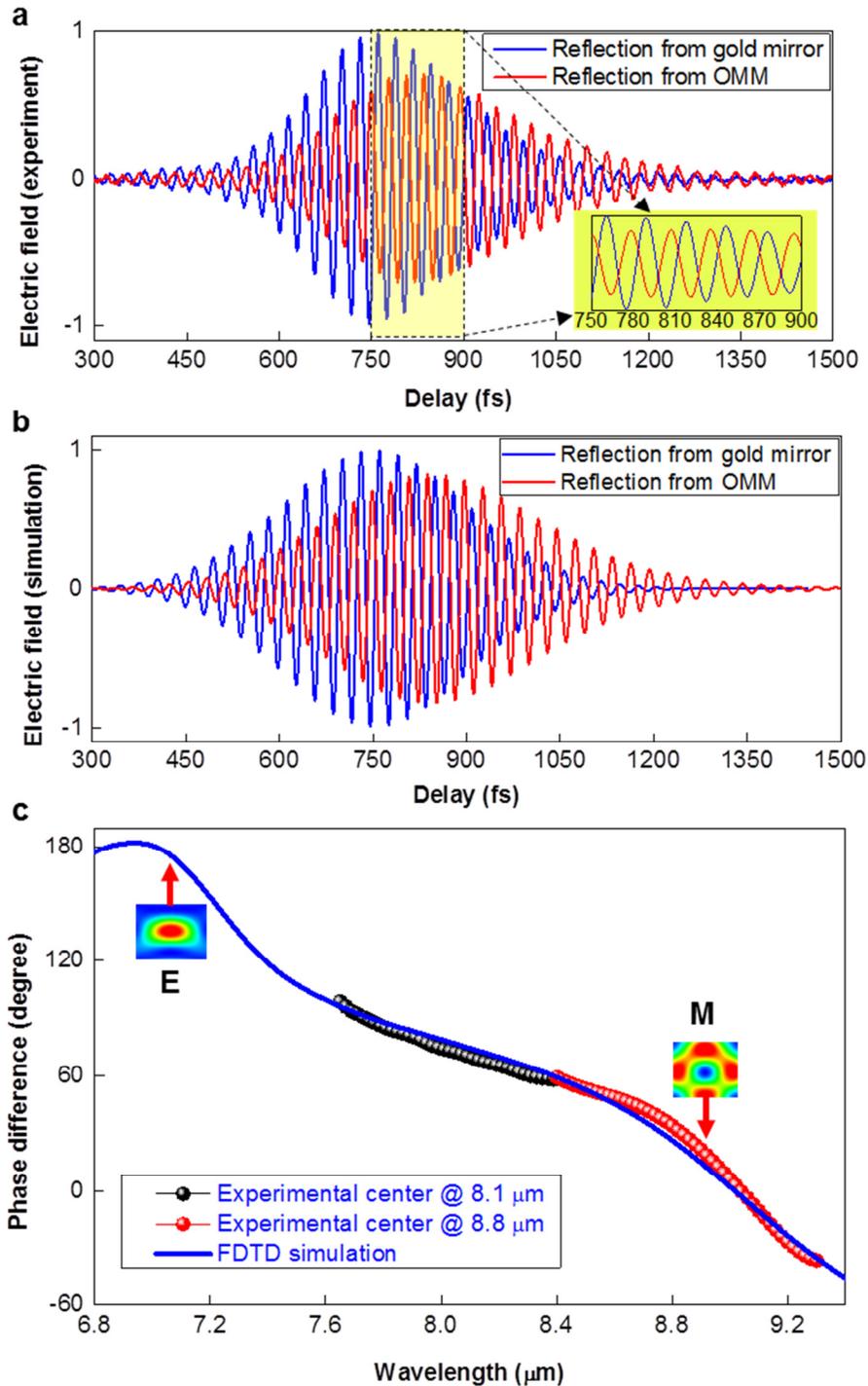

**Figure 3 | Electric field transients measured by time domain spectroscopy. a**, Experimental measurement of the electric field reflected from the gold surface (blue curve) and OMM (red curve) at the OMM magnetic resonance. The gold surface serves as a reference. **b**, FDTD simulation of the reflected electric field from the gold surface (blue curve) and the OMM (red curve). Both experiment and

simulation show that the electric field fringes from the OMM are out-of-phase with those from the gold surface (i.e. a normal mirror). This unambiguously demonstrates the magnetic mirror behavior of the OMM. **c**, Experimental and FDTD simulation results of the optical phase of the reflected wave. To cover a broader spectral range, two experimental data sets were obtained with the central frequency tuned to 8.8 µm and 8.1 µm.

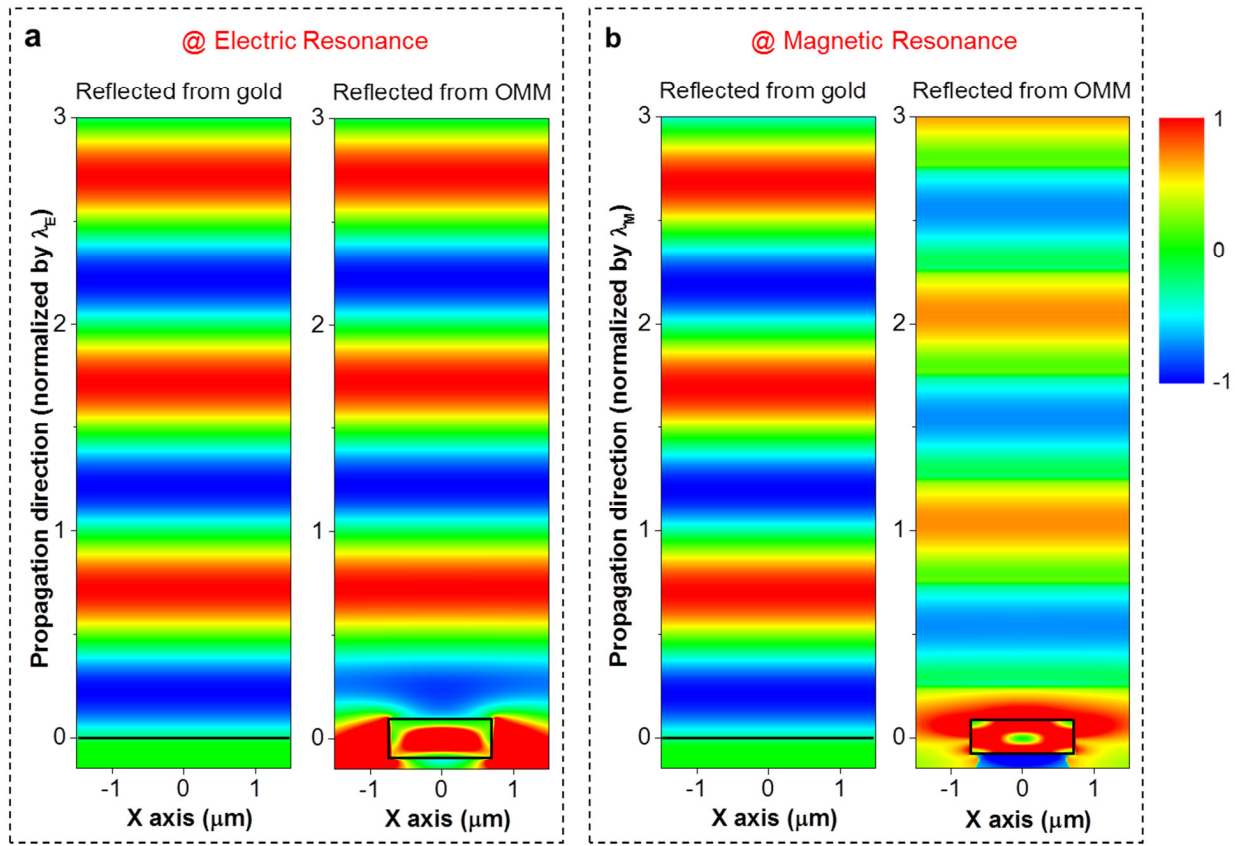

**Figure 4 | Standing wave patterns of light reflected from a gold mirror and an OMM at (a) the OMM electric resonance and (b) the OMM magnetic resonance.** The OMM behaves like a conventional mirror at the electric resonance shown in **a** and behaves like a magnetic mirror at the magnetic resonance shown in **b**. The black lines and rectangles represent the gold-air interfaces and the boundaries of the Te cubic resonators, respectively. A node of the standing wave always occurs at the gold-air interface. On the contrary, the top of the cubic resonator is at the node for the electric resonance and at the antinode for the magnetic resonance. All patterns share the same color scale bar on the right. Also note that the aspect ratio used for this figure causes the profile of the cubic resonator to appear as a rectangle.

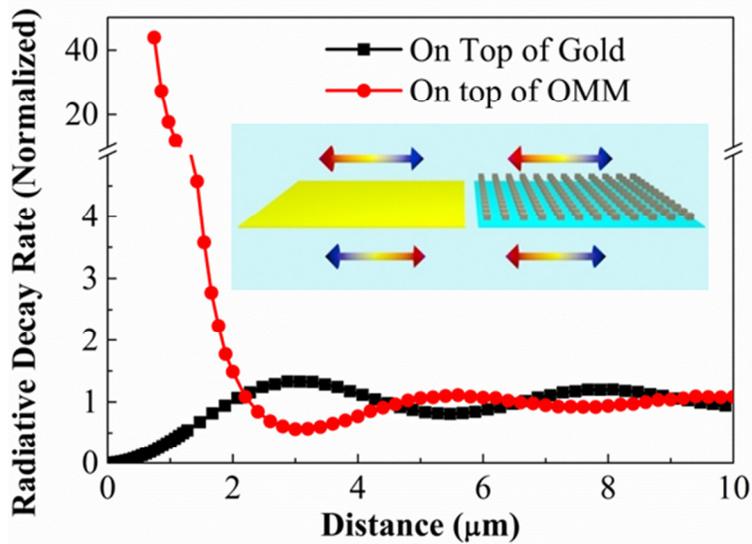

**Figure 5 | Radiative decay rate of a transverse electric dipole near gold and OMM surfaces**. The normalized radiative decay rate of a transverse electric dipole oscillating at the magnetic resonance frequency as a function of the dipole-surface separation for both a 5 x 5 array approximating our OMM (red curve) and a gold surface (black curve). The distance of the dipole from the OMM is calculated using the center of the cubic resonator as distance "0" which is in agreement with our theoretical calculation (Supplementary Information 4). First, the oscillatory dependence on distance is shifted by about half a period. Second, while the emission from the dipole is quenched very close to the gold surface, the dipole emission near the magnetic mirror is enhanced even for very small distances. (Inset: Schematic of an electric dipole placed on top of a typical mirror and its reversed image dipole (Left). Schematic of an electric dipole on top of a dielectric magnetic mirror and its un-reversed image dipole (Right) at the magnetic dipole resonance.)

**Acknowledgments**

We thank Larry K. Warne for useful technical discussions. This work was performed, in part, at the Center for Integrated Nanotechnologies, a U.S. Department of Energy, Office of Basic Energy Sciences user facility. Portions of this work were supported by the Laboratory Directed Research and Development program at Sandia National Laboratories. Sandia National Laboratories is a multi-program laboratory managed and operated by Sandia Corporation, a wholly owned subsidiary of Lockheed Martin Corporation, for the U.S. Department of Energy's National Nuclear Security Administration under contract DE-AC04-94AL85000.

# Supplementary Information

## S1- Spectral locations of the magnetic and electric resonances.

Transmission, reflection and loss of an OMM

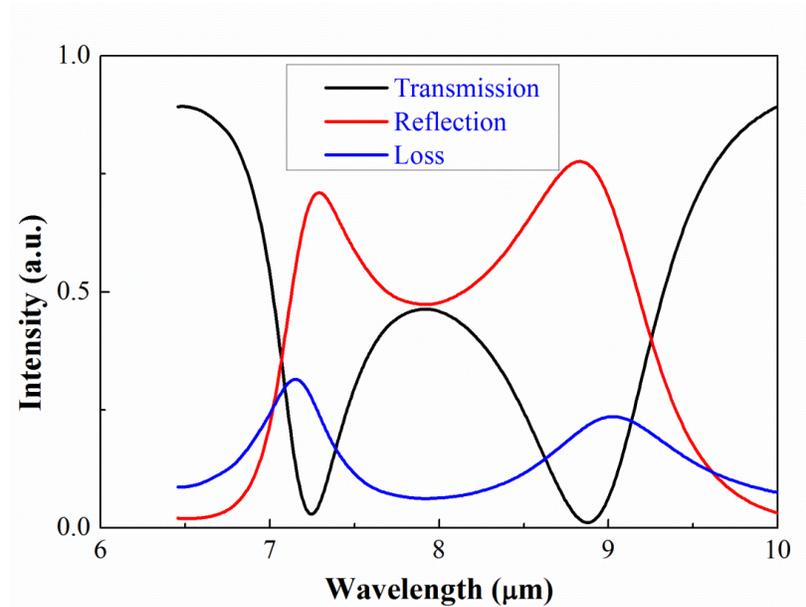

**Figure S1-1 |** Simulated transmission, reflection and loss spectra of an OMM. Loss is calculated as: Loss=1−transmission−reflection[1].

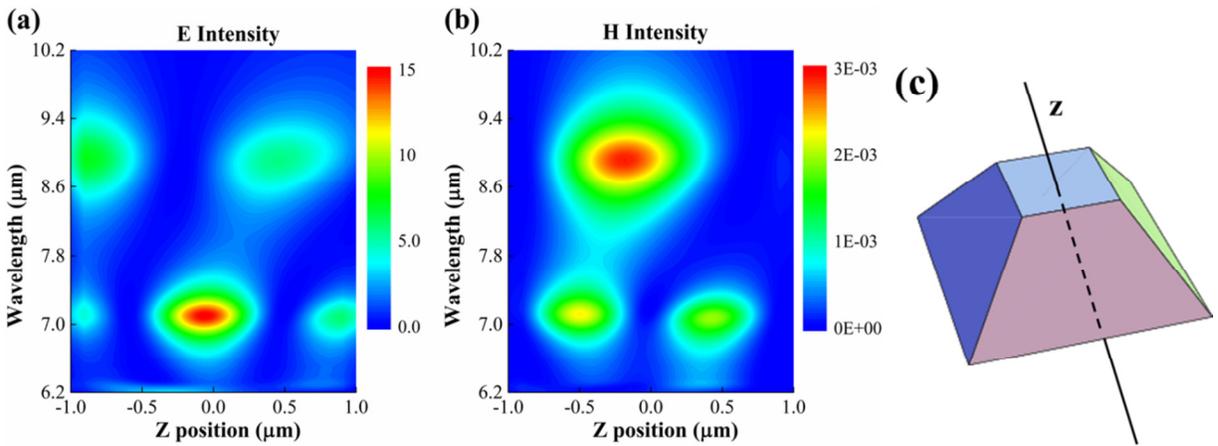

**Figure S1-2 |** Electric intensity (**a**) and magnetic intensity (**b**) as a function of wavelength and location inside the cube (along the z axis shown in (**c**)) showing the strongest electric and magnetic intensity located at z=-0.06 µm and -0.18 µm (z is the axis passing through the center of the truncated pyramid shape resonator, z=0 is the height center). Therefore, we locate the effective reflection plane near the

center of the cube. The intensity maxima are located slightly below this center plane because of the substrate and the truncated pyramid shape. Simulations show that both electric and magnetic field maxima are located at the center when the resonator geometry is that of a perfect cubic and no substrate is present. This figure also shows the strongest electric and magnetic dipole resonances at λ=8.95 µm and 7.08 µm.

**S2- Three ways of confirming that switching between the gold surface and the OMM did not cause spurious delay change**

The reflection time-domain-spectroscopy system was aligned to make sure no spurious delay change occurs when the reflection spot is switched from the OMM to the gold surface. As illustrated in Figure S2, the red dots labeled "1", "2" and "3" represent three equally spaced focal spots of the fs mid-IR beam on the OMM and the gold surface. Before translating between spots "1" and "2", we first move the beam between spots "2" and "3" (both on gold). If the sample is not perpendicular to the plane of incidence, translation of the sample will cause a spurious delay change. We developed three methods for correcting this spurious delay change: 1) Adjusting the sample's angle until TDS signals of the reflected electric fields from "2" and "3" temporally overlap with each other; 2) Using the spurious delay change between "2" and "3" to correct the translation between "1" and "2"; and 3) Using the reflected electric field phase difference (calculated using the Fourier transform) between spots "2" and "3" to calculate the spurious delay change.

Moreover, our TDS experiment used a digital oscilloscope which records the signal at $10^3$—$10^6$ delay positions and averaged hundreds of scans as the delay stage moved back and forth. Several hundreds of scans can be averaged in 2-3 minutes.

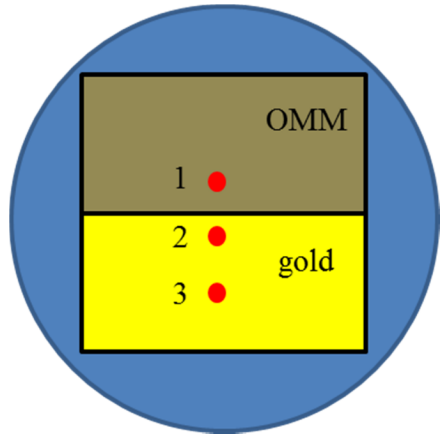

**Figure S2 |** Schematic showing the positions of the mid-IR focal spots used to correct the spurious delay changes caused by translation.

**S3- Mid-IR time domain spectroscopy setup in detail.**

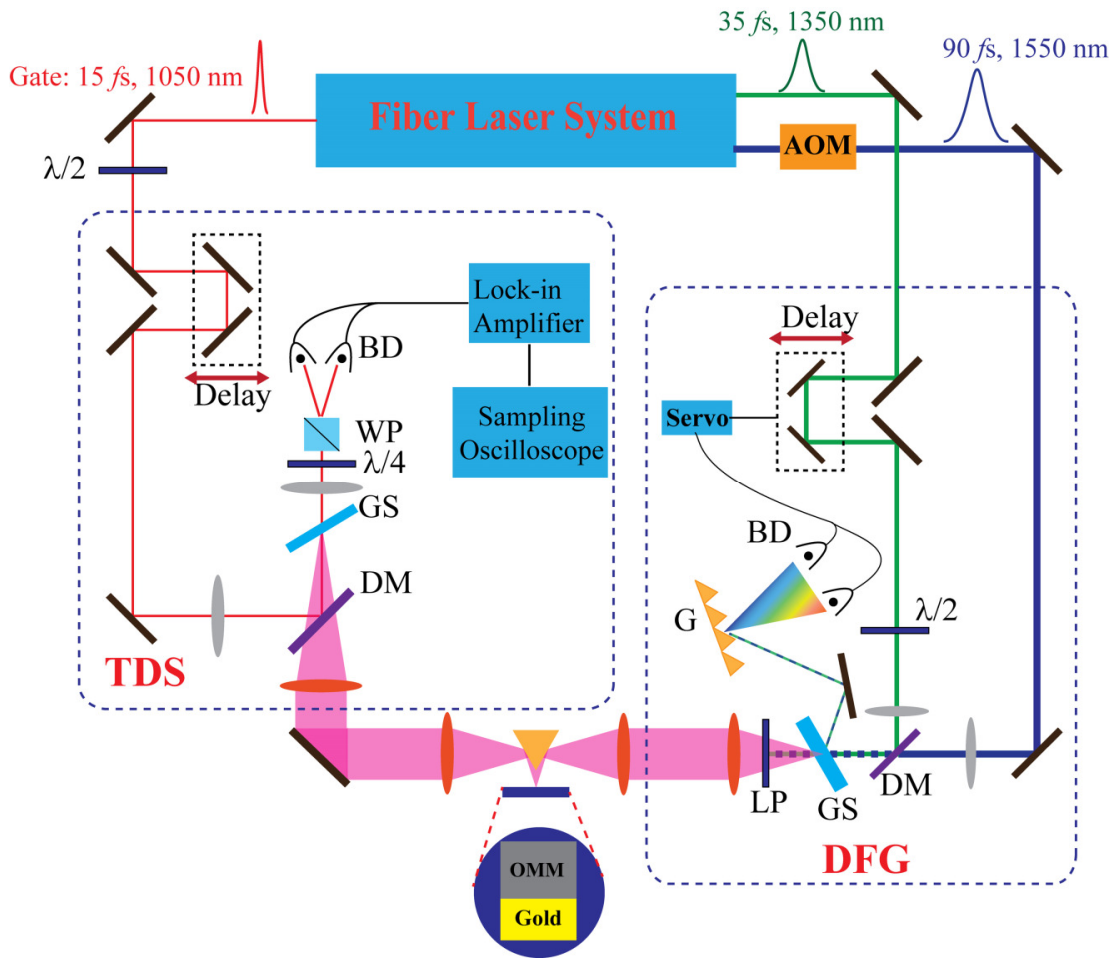

**Figure S3 | Schematic of the difference frequency generation (DFG) and time-domain spectroscopy (TDS) set-up[2].** We used a phase-locked TDS in the mid-IR to directly measure the phase shift of the reflected electric field from the OMM. Mid-IR pulses of 250 fs duration, produced by difference frequency mixing between ~1.35 µm and ~1.55 µm pulses, were focused by a ZnSe lens onto the OMM or gold surface. Another ZnSe lens was used to collect the reflected mid-IR beam. Gate pulses with a wavelength of 1.05 µm and duration of 15 fs were combined with the mid-IR pulses using a dichroic mirror. Both the gate and the mid-IR pulses were focused into a 0.5 mm thick GaSe crystal, taking care to match the focal widths of the beams. The two polarizations are separated using a quarter wave plate and a Wollaston prism, and detected by a balanced detector. AOM, acousto-optic modulator; DM, dichroic mirror; GS, GaSe nonlinear crystal; LP, long-wavelength pass filter; G, diffraction grating; BD, balanced detector; $\lambda/2$, half-wave plate; $\lambda/4$, quarter-wave plate; WP, Wollaston prism.

**S4-Theoretical explanation of the observed reflection phases employing two-dimensional periodic dyadic Green's functions.**

In this section we provide a theoretical basis for the expected phase difference between the magnetic mirror and the electric mirror behaviors of the OMM, and show that the "proper" location for the placement of the reference mirror is the center of the cubes. For convenience, we consider perfect cubic shape resonators with a side dimension of 1.45 µm in the analyses performed in this section. The electric and magnetic polarizabilities of an isolated cube (possessing the experimentally determined permittivity of Te) are computed following the procedure described in Ref. 3-4 and are shown in Fig. S4-1. Similar results have been recently reported in Ref 5. In Fig. S4-1, electric and magnetic resonances are clearly visible at wavelengths of about 6.9 µm and 9 µm, respectively, in agreement with what was reported in Fig. 1c.

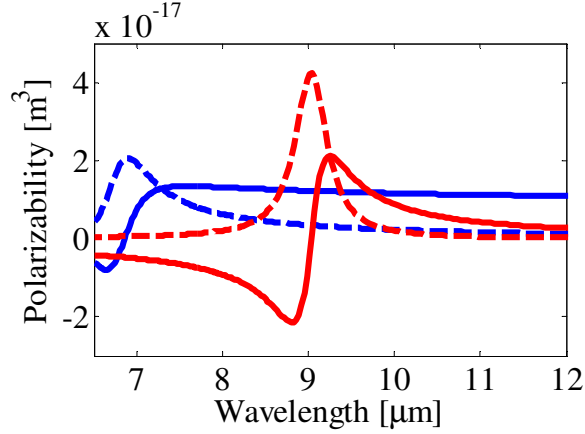

**Figure S4-1 |** Electric (blue) and magnetic (red) cube polarizabilities. Solid lines are real parts, dashed lines imaginary parts.

According to the result in Fig. S4-1, in the vicinity of the electric resonance we can approximate the scattering response of each cube as an electric dipole response. Similarly, near the magnetic resonance, we can approximate the scattering response of each cube as a magnetic dipole response.

We now consider a 2D array of cubes (assuming for simplicity the resonators are in free space without a substrate). Using the polarizabilities in Fig. S4-1, we can build the two independent systems of equations describing an array of either electric or magnetic dipoles only as (a more complete formulation of the problem that includes coupling between the electric and magnetic dipoles will be given below)

$$\left(\underline{\mathbf{I}} - \alpha_{ee}\underline{\breve{\mathbf{G}}}_{ee}^{\infty}\right)\cdot\mathbf{p} = \alpha_{ee}\mathbf{E}^{inc} \qquad \left(\underline{\mathbf{I}} - \alpha_{mm}\underline{\breve{\mathbf{G}}}_{mm}^{\infty}\right)\cdot\mathbf{m} = \alpha_{mm}\mathbf{H}^{inc} \qquad \text{(S-Eq1)}$$

where **p** and **m** are the induced electric and magnetic dipole moments of the reference dipoles in the array, $\alpha_{ee}$ and $\alpha_{mm}$ are the cube's electric and magnetic polarizabilities reported in Fig. S4-1, $\underline{\breve{\mathbf{G}}}_{ee}^{\infty}$ and $\underline{\breve{\mathbf{G}}}_{mm}^{\infty}$ are the regularized periodic dyadic Green's functions for the phased array of electric and magnetic dipoles[6,7], $\mathbf{E}^{inc}$ and $\mathbf{H}^{inc}$ are the incident electric and magnetic fields, and $\underline{\mathbf{I}}$ is the unit dyad. By solving the two systems in Eq. (1), we are able to evaluate **p** and **m** including field interactions among all the (electric or magnetic) dipoles in the array. The knowledge of **p** and **m** is essential for the estimation of reflection coefficients and scattered fields. In particular, limiting the analysis only to the propagating fundamental plane wave with longitudinal wavenumber $k_z = k_0$ radiated by the dipoles, with $k_0$ the free space wavenumber, an array of magnetic dipoles radiates a magnetic field as

$$\mathbf{H}_m = \underline{\mathbf{G}}_{mm} \cdot \mathbf{m}, \quad \underline{\mathbf{G}}_{mm} = \frac{e^{ik_z(z-z_0)}}{-2iabk_z}\left[k_0^2 \underline{\mathbf{I}} - \mathbf{kk}\right] \quad \text{(S-Eq2)}$$

whereas an array of electric dipoles radiates an electric field as

$$\mathbf{E}_e = \underline{\mathbf{G}}_{ee} \cdot \mathbf{p}, \quad \underline{\mathbf{G}}_{ee} = \frac{e^{ik_z(z-z_0)}}{-2iab\varepsilon_0 k_z}\left[k_0^2 \underline{\mathbf{I}} - \mathbf{kk}\right]. \quad \text{(S-Eq3)}$$

In Eqs. (2) and (3), $a$ and $b$ indicate the periods in the $x$ and $y$ directions (equal to 3.45 µm in the full-wave simulation), $\mathbf{k} = k_x\hat{\mathbf{x}} + k_y\hat{\mathbf{y}} \pm k_z\hat{\mathbf{z}}$ is the wavevector (+/− is used when the observation point $z$ is above/below the source point $z_0$), and **p** and **m** are the induced electric and magnetic dipole moments calculated solving Eq. (1). The dyadic Green's functions $\underline{\mathbf{G}}_{ee}$ and $\underline{\mathbf{G}}_{mm}$ in Eqs. (2) and (3) are computed using a spectral approach[7,8] and model only the fundamental propagating plane wave — the only wave responsible for specular reflection. In accordance with the experiment, we further assume normal plane wave excitation with $\mathbf{E}^{inc} = E_0 e^{-ik_0 d_0}\hat{\mathbf{y}}$, where $z = d_0$ represents the zero-phase distance from the array plane $z_0 = 0$. For comparison with the experimental measurements, we define the *electric field* reflection coefficient at the electric and magnetic resonances as $\Gamma_e$ and $\Gamma_m$ respectively. These reflection coefficients are defined as the ratio of the transverse electric field components at the array plane and are given by

$$\Gamma_e = \frac{\mathbf{E}_e \cdot \hat{\mathbf{y}}}{\mathbf{E}^{inc} \cdot \hat{\mathbf{y}}} \propto -\frac{\alpha_{ee}}{i} \qquad \Gamma_m = -\frac{\mathbf{H}_m \cdot \hat{\mathbf{x}}}{\mathbf{H}^{inc} \cdot \hat{\mathbf{x}}} \propto \frac{\alpha_{mm}}{i} \quad \text{(S-Eq4)}$$

Note that for $\Gamma_m$ we have expressed the electric field reflection coefficient in terms of the incident and reflected magnetic fields, which requires the inclusion of the minus sign in this expression. Closer inspection of Eq. (4) shows the phases of both $\alpha_{ee}$ and $\alpha_{mm}$ at resonance are 90 degrees, and are compensated by the imaginary term at the denominator of the two expressions in Eq. (4). Therefore, the 180 and 0 degrees phase shifts for $\Gamma_e$ and $\Gamma_m$ for the respective resonances inherently originate from the presence and absence of a minus sign in their corresponding equations in Eq. (4). In general, there could be an additional phase shift in these expressions due to the $\left(\underline{\mathbf{I}} - \alpha_{ee} \underline{\breve{\mathbf{G}}}_{ee}^{\infty}\right)^{-1}$ and $\left(\underline{\mathbf{I}} - \alpha_{mm} \underline{\breve{\mathbf{G}}}_{mm}^{\infty}\right)^{-1}$ factors in the solution of Eq. (1), however this phase shift is expected to be small. A plot of the two reflection coefficients evaluated at the array plane $z = z_0$ is reported in Fig. S4-2 (this plane is assumed to be at the cubes' center). We observe that the phase of the reflection coefficient $\Gamma_e$ at the electric resonance frequency is about 180 degrees, signature of a perfect electric conductor behavior. We further observe about 0 degrees phase for the reflection coefficient $\Gamma_m$ at the magnetic resonance frequency, signature of a perfect magnetic conductor behavior. This result is in agreement with the experimental observations of the OMM behavior in the main text of this paper.

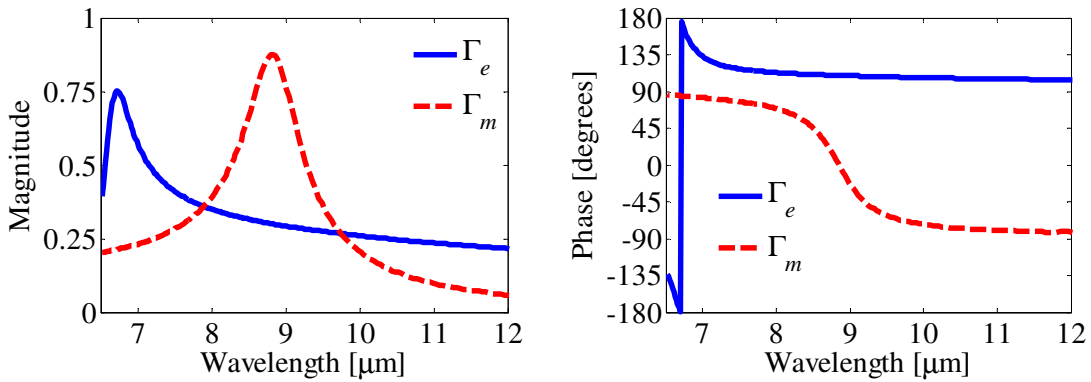

**Figure S4-2 |** Plots of the two reflection coefficients from the array of cubes modeled as an array of either electric (solid blue) or magnetic (dashed red) dipoles.

For resonators that are very small compared to the operating wavelength it is obvious that the correct location of the reference mirror should be at the center of the resonators. This assumption has been implicitly made in the dipole formulations given earlier in this section. However, as the resonators become larger it is not clear whether the incident field will experience any additional phase delay as it propagates from the top surface to the center of the array of high permittivity cubes. While the neglect of coupling between the electric and magnetic dipoles was allowable when considering the relative phase shift between the reflected electric field at the two resonances, such a description cannot be used to determine the correct location of the reference mirror to which the experimental phases are referenced, nor for a more accurate estimation of the (total) electric field reflection coefficient measured experimentally. Such an extra phase shift might require that the mirror reference plane be displaced from the cubes' center. To determine whether such an additional phase shift is required, we compute the scattered electric and magnetic fields at $z = d_0$ (i.e., at the location of the plane wave injection in full-wave simulations, considered to be in the far-field of the metasurface with $d_0 = 9.275$ µm ) using both a dipolar method and full-wave simulations. For the dipolar method, we now consider coupled electric and magnetic dipoles adapting the formulation in Ref. 5&9 to the case of 2D periodic arrays, as for example recently shown in Ref. 10. In other words, we find **p** and **m** by solving the 6 by 6 system

$$\left[\begin{pmatrix} \boldsymbol{\alpha}_{ee}^{-1} & \mathbf{0} \\ \mathbf{0} & \boldsymbol{\alpha}_{mm}^{-1} \end{pmatrix} - \begin{pmatrix} \breve{\underline{\mathbf{G}}}_{ee}^{\infty} & \breve{\underline{\mathbf{G}}}_{em}^{\infty} \\ \breve{\underline{\mathbf{G}}}_{me}^{\infty} & \breve{\underline{\mathbf{G}}}_{mm}^{\infty} \end{pmatrix}\right] \cdot \begin{pmatrix} \mathbf{p} \\ \mathbf{m} \end{pmatrix} = \begin{pmatrix} \mathbf{E}^{inc} \\ \mathbf{H}^{inc} \end{pmatrix} \quad \text{(S-Eq5)}$$

where the dyads $\breve{\underline{\mathbf{G}}}_{em}^{\infty}$ and $\breve{\underline{\mathbf{G}}}_{me}^{\infty}$ introduce electromagnetic coupling between different dipoles[5,9,10];. In full-wave simulations, we consider a 2D periodic array of cubes in free space under normal incidence. Fig. S4-3 shows the magnitude and phase of the reflection coefficient referred at the array plane using both the dipole theory and the full-wave simulations. We observe a remarkable agreement between the theoretical and full-wave results over the entire frequency range, given the approximate nature of the dipolar methods (discrepancies may still be due to multipolar contributions and/or spatial dispersion). In particular, when looking at the phase result, we do not observe a significant additional phase shift. This result confirms that any extra phase shift due to the presence of the real cubes is small and the approximation of locating the mirror reference plane at the cubes' center is a proper one. For comparison, we also plot the experimentally determined reflection phase in Fig. S4-3. Surprisingly good agreement with the theory and simulation is observed, considering that the theory and simulation utilized perfect cubes rather than the truncated pyramids measured in the experiment, and since the theory and simulation did not include the effect of the substrate.

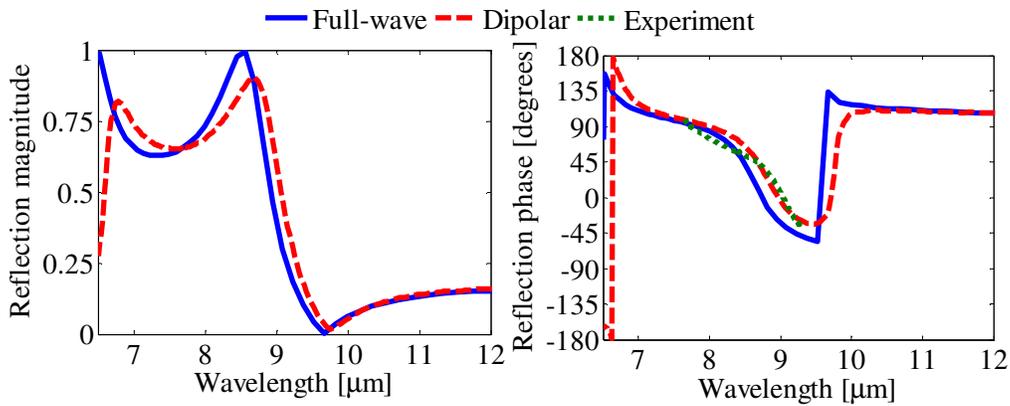

**Figure S4-3 |** Magnitude and phase of the reflection coefficient at the array plane computed with full-wave simulations (blue solid) and dual (electric and magnetic) dipolar approximation (red dashed). The experimental result is reported for completeness as a green dotted curve.

**S5-Electric field fringes reflected from a gold mirror and the OMM closely match each other at the electric resonance wavelength of the OMM**

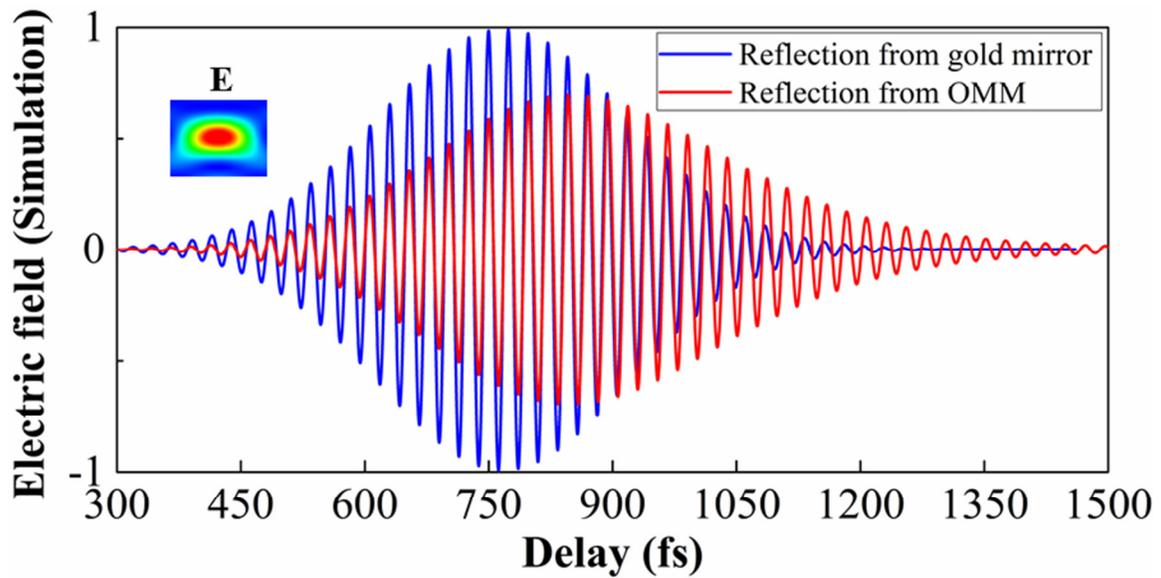

**Figure S5 |** Simulated electric field transients reflected from a gold mirror (blue curve) and the OMM (red curve) at the electric dipole resonance.

**S6-Phase different between magnetic and electric resonances for another OMM sample**

We also measured the phase of the reflected electric field using our TDS system for a different OMM sample which exhibited a smaller spectral separation between the electric and magnetic resonances compared with the OMM shown in Fig. 3. The experimental results show that within a narrow wavelength range of electric and magnetic resonances, the phase difference between the resonances is almost the same as for the sample described in the main text.

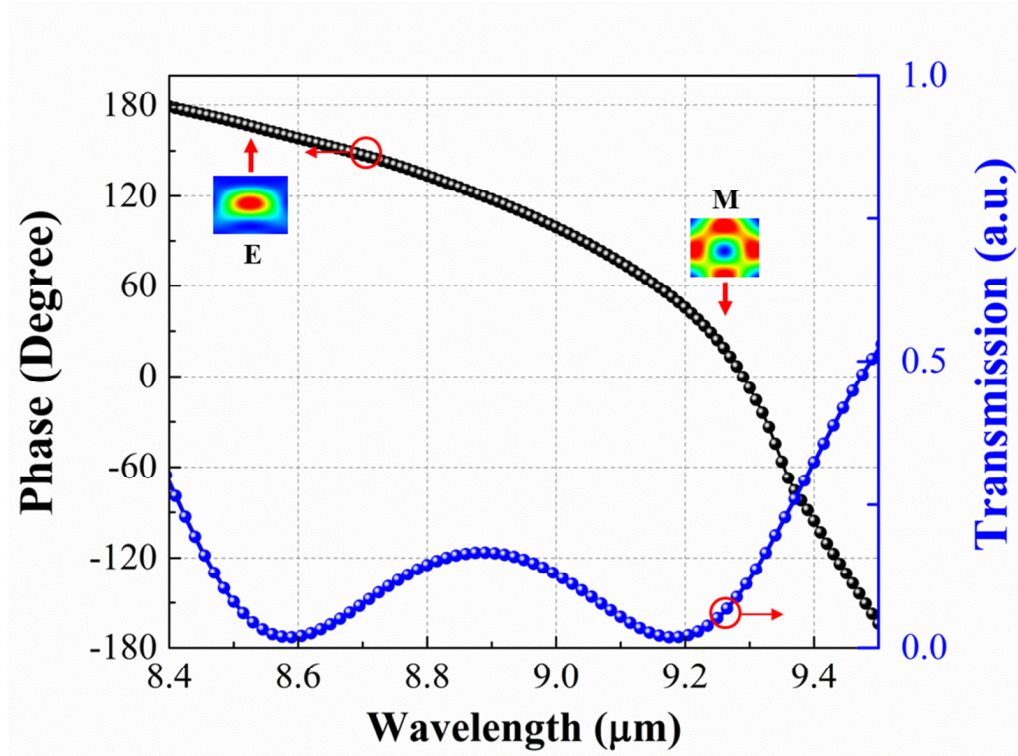

**Figure S6 |** Experimental results of another OMM sample with smaller spectral difference between electric and magnetic dipole resonances. Therefore, our mid-IR fs pulses are spectrally broad enough to cover both resonances without tuning. Despite the difference between this OMM sample and the sample discussed in the main text, we observe close to 180 degree phase difference between electric and magnetic resonances in both cases. This unambiguously demonstrates an OMM formed by Te dielectric resonators at magnetic dipole resonance.